\documentclass[prb,twocolumn,showkeys,showpacs]{revtex4}

\usepackage{amsmath}
\usepackage{graphicx}

\begin{document}

\title{Low conductance of the nickel atomic junctions in hydrogen atmosphere}

\author{Shuaishuai Li}
\affiliation{Department of Physics, Shanghai Normal University, Shanghai, 200234, P.R. China}

\author{Yi-Qun Xie}
\email{yqxie@shnu.edu.cn}
\affiliation{Department of Physics, Shanghai Normal University, Shanghai, 200234, P.R. China}

\author{Yibin Hu}
\email{ybhu@mail.sitp.ac.cn}
\affiliation{National Laboratory for Infrared Physics, Shanghai Institute of Technical Physics,  Chinese Academy of Sciences, Shanghai, 200083, P.R. China}

\begin{abstract}
The low conductance of nickel atomic junctions in the hydrogen environment is studied using the nonequilibrium Green's function theory
combined with first-principles calculations. The Ni junction bridged by a $H_2$ molecule has a conductance of approximately 0.7 $G_0$.
This conductance is contributed by the anti-bonding state of the $H_2$ molecule, which forms a bonding state with the $3d$ orbitals
of the nearby Ni atoms. In contrast, the Ni junction bridged by the two single H atoms has a conductance of approximately 1 $G_0$,
which is weakly spin-polarized. The spin-up channels were found to contribute mostly to the conductance at a small junction gap,
while the spin-down channels play a dominant role at a larger junction gap.
\end{abstract}



\date{\today}

\maketitle

\section{Introduction}

Electronic transport properties of metal atomic junctions (MAJs) \cite{nature-395-780,prl-87-096803,prl-91-096801,prb-69-081401,prb-69-125409,prb-70-113107,prb-73-125424,prb-73-075405,jpcc-7-9903,prb-81-125422,prb-82-075417,jpcl-1-923,prl-94-036807,prl-98-146802,prb-84-245412,prb-87-075415,njp-10-033005,prl-107-126802,prb-89-075436,prb-91-245404,jap-117-043902,apl-101-192408,jap-117-064310,fp-9-780}. have been extensively studied in recent years because of their potential applications in electronic nanodevices. MAJs have shown various electronic transport behaviors such as the even-odd conductance oscillations \cite{prl-87-096803,prb-70-113107}, high magnetoresistance ratio \cite{prb-89-075436}, and spin-filtering effects \cite{jap-117-043902}. Of particular interest is the transport behavior of the MAJs in an atmosphere of a gas such as hydrogen, oxygen, or CO. With the adsorption of gas molecules or atoms, the mechanical stability of the MAJs can be largely improved to generate various atomic configurations with special conductance. For example, in the hydrogen environment, a low conductance of approximately 1 $G_0$ has been observed for the nonmagnetic junctions of Rh \cite{prb-81-125422}, Pb \cite{prb-82-075417}, and Pt \cite{prl-94-036807}, as well as for magnetic ones such as Fe and Co \cite{prb-69-081401,jpcl-1-923} junctions. Here $G_0 = 2e^2/h$, where $e$ is electron charge and $h$ is Planck's constant. Much effort has also been put into understanding the physical mechanisms of such low conductance. For instance, it has been proposed that the 1 $G_0$ conductance of the Pt junction is facilitated by the anti-bonding state of the hydrogen molecule in a $Pt-H_2-Pt$ configuration \cite{prl-94-036807}.  In contrast, the origin of the 1 $G_0$ conductance for the Co junction in the hydrogen environment can be attributed to the adsorption of the single H atoms \cite{jap-117-064310}.

In a recent experiment, a conductance of 1 $G_0$ was experimentally observed for the nickel junction in the deuterium (D2) gas atmosphere \cite{prb-91-245404}. However, the underlying mechanism of the 1 $G_0$ conductance remains to be explored. This work studies the spin-polarized transport properties of the Ni atomic junction bridged by a $H_2$ molecule or single H atoms, to give an insight into the physical origin of this 1 $G_0$ conductance. Note that since $H_2$ and $D_2$ have essentially the same electronic structure, they also possess similar electronic transport properties.  

\section{Modeling and methods}

\begin{figure}[h]
\begin{center}
\includegraphics[scale=0.36]{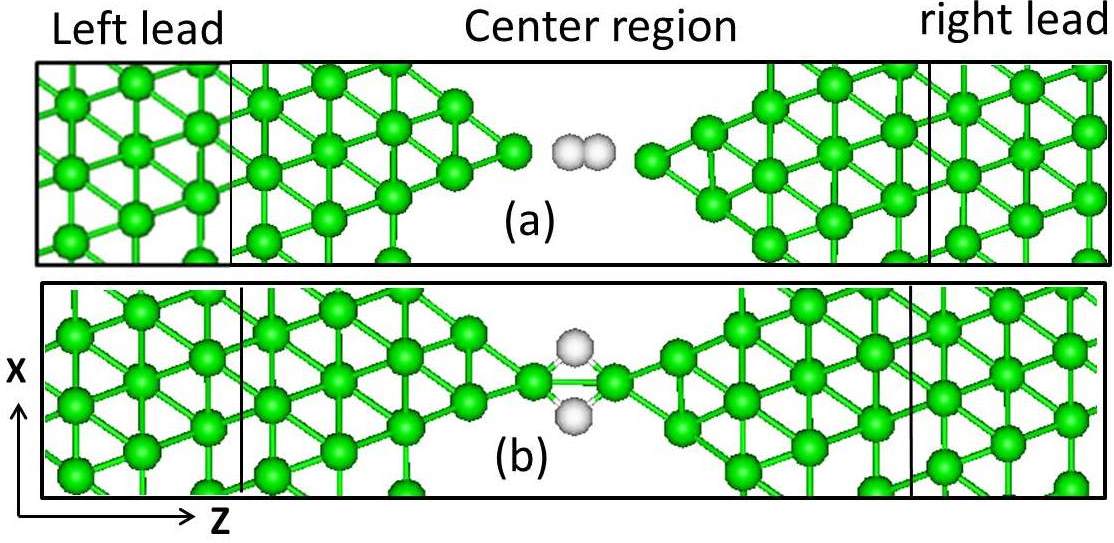}
\caption{\label{fig1}
Schematic diagram of the Ni atomic junction (a) bridged by a $H_2$ molecule ($Ni-H_2-Ni$ junction), and (b) by the two single H atoms absorbed symmetrically between the two Ni tip atoms ($Ni-2H-Ni$ junction). Green spheres denote Ni atoms, and others are H atoms. The electrical transport is along the Z direction.}
\end{center}
\end{figure}

Two junction structures were considered first, and their conductance values were calculated. As shown in Fig. \ref{fig1}, the Ni atomic junction is mimicked by a two-probe system consisting of the left and right leads and a center region. Each lead comprises three Ni (111) layers with ($3 \times 3$) periodicity in the $x-y$ plane, and extends to $z = \pm\infty $. In the center region, there are two pyramidal tips connected by an $H_2$ molecule (Fig. \ref{fig1}(a)) or by the two single H atoms adsorbed symmetrically between the two tip atoms (Fig. \ref{fig1}(b)). These two configurations are therefore named the $Ni-H_2-Ni$ junction and $Ni-2H-Ni$ junction, respectively. The reason for choosing such configurations is that they can account for the 1 $G_0$ conductance of several MAJs in the hydrogen atmosphere, according to the previous study \cite{prl-94-036807,jap-117-064310}.

The center region is relaxed using the VASP code \cite{prb-54-11169}. During the relaxation, the hydrogen atoms and the tip atoms were allowed to be fully relaxed until all the residual force on atoms were smaller than 0.03 eV/\AA, while the other atoms were fixed. In experiments, recording the conductance during the stretching process of the junction identifies the conductance of the atomic junction. To mimic the stretching process, the distance between the two tips was increased in steps of 0.2 \AA, and the conductance of the junction was calculated at each step.   

For the VASP calculation, the projector augmented-wave method \cite{prb-50-17953} was used for the wave function expansion with an energy cutoff of 400 eV. The PW91 version \cite{prb-46-6671} of the generalized gradient approximation was adopted for the electron exchange correlation. The Brillouin zone was sampled using a $4 \times 4 \times 1$ grid of the Monkhorst-Pack k points \cite{prb-13-5188}. In transport calculations, the first-principles nonequilibrium Green¡¯s function (NEGF) approach \cite{prb-70-085410,prb-63-245407} was adopted. A numerical double zeta plus polarization basis set (DZP) was used for the wave function expansion. The transmission spectrum was calculated with a $8 \times 8 \times 1$ grid of the Monkhorst-Pack k points. In the following, all conductance values are presented in units of $G_0 = 2e^2/h$.

\section{Results and discussion}

\begin{figure}[b]
\begin{center}
\includegraphics[scale=0.3]{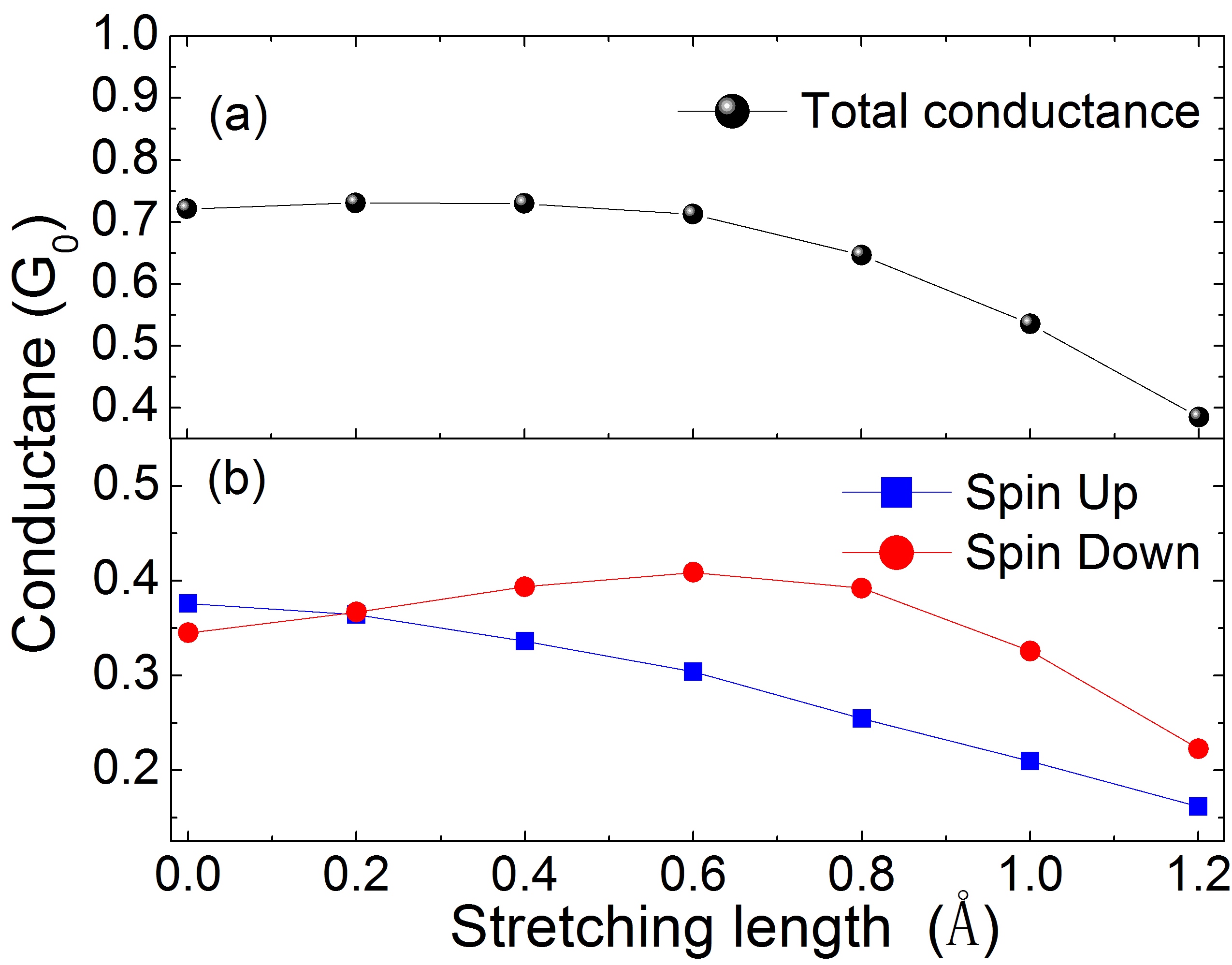}
\caption{\label{fig2}
(a) the total conductance and (b) the spin-polarized conductance of the $Ni-H_2-Ni$ junction during the stretching process.}
\end{center}
\end{figure}

First, the conductance of the $Ni-H_2-Ni$ junction is investigated.  For the initially relaxed configuration in Fig. \ref{fig1}(a), an H-H bond of 0.88 \AA, which is about 0.1 \AA greater than that of the free $H_2$ molecule (indicating that the H molecule experiences an attractive force from the nearby Ni tip atoms) is obtained. When the junction is stretched along the $z$ direction, the H-H bond is shortened to 0.76 \AA, owing to the weakened attraction from the nearby Ni tips. During the stretching process, the conductance of the junction is calculated and is presented in Fig. \ref{fig2}(a). It was found that the conductance retains a stable value of approximately 0.7 $G_0$ for the stretching length from 0 to 0.6 \AA, after which it decreases sharply as the junction is elongated further. Moreover, the conductance is spin-polarized. As shown in Fig. \ref{fig2}(b), the spin-up conductance is greater than spin-down conductance for the non-stretched junction, while the spin-down component contributes mostly to the total conductance during the stretching process. 

\begin{figure}[h]
\begin{center}
\includegraphics[scale=0.75]{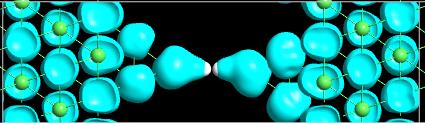}
\caption{\label{fig3}
The local density of states of 0.02 states/($\text{\AA}^3 \cdot \mathrm{eV}$) for the $Ni-H_2-Ni$ junction at the stretch length of 0.2 \AA.}
\end{center}
\end{figure}

To understand the electronic origin of the conductance, the local density of states of the junction at the stretching length of 0.2 \AA, as shown in Fig. \ref{fig3}, was calculated. It can be seen that the $H_2$ molecule possesses an anti-bonding state that forms a strong bond with the $3d$ orbitals of the nearby Ni atoms. This indicates that the transport of this nickel junction is conveyed through the anti-bonding state of the $H_2$ molecule other than its bonding state.

\begin{figure}[h]
\begin{center}
\includegraphics[scale=0.6]{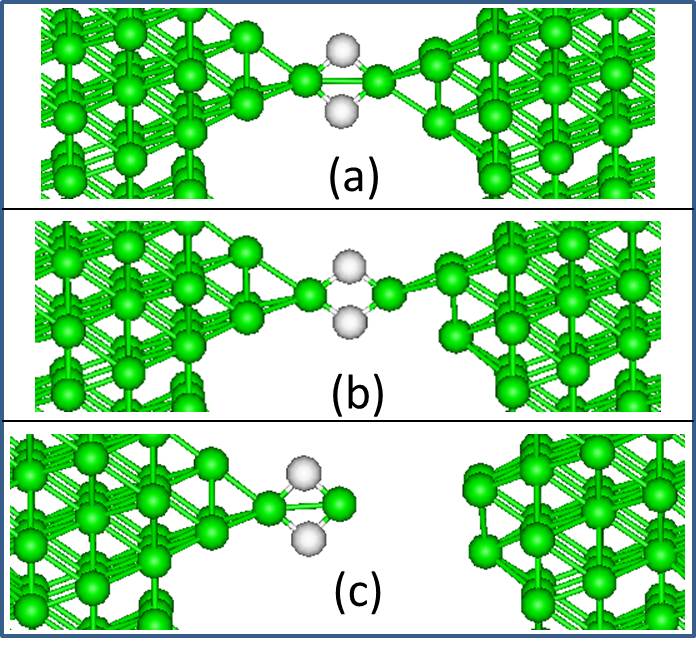}
\caption{\label{fig4}
Structure evolution of the $Ni-2H-Ni$ junction during the stretching process.}
\end{center}
\end{figure}

The Ni-2H-Ni junction shown in Fig. \ref{fig1}(b) is considered next. The evolution of this junction during the stretching process is shown in Fig. \ref{fig4}. The tips suffer an evident deformation when the junction is stretched (see Figs. \ref{fig4}(a)$\rightarrow$(b)), and finally the junction is broken as the tip-apex atom is pulled off from the right tip (Fig. \ref{fig4}(c)).  

\begin{figure}[h]
\begin{center}
\includegraphics[scale=0.4]{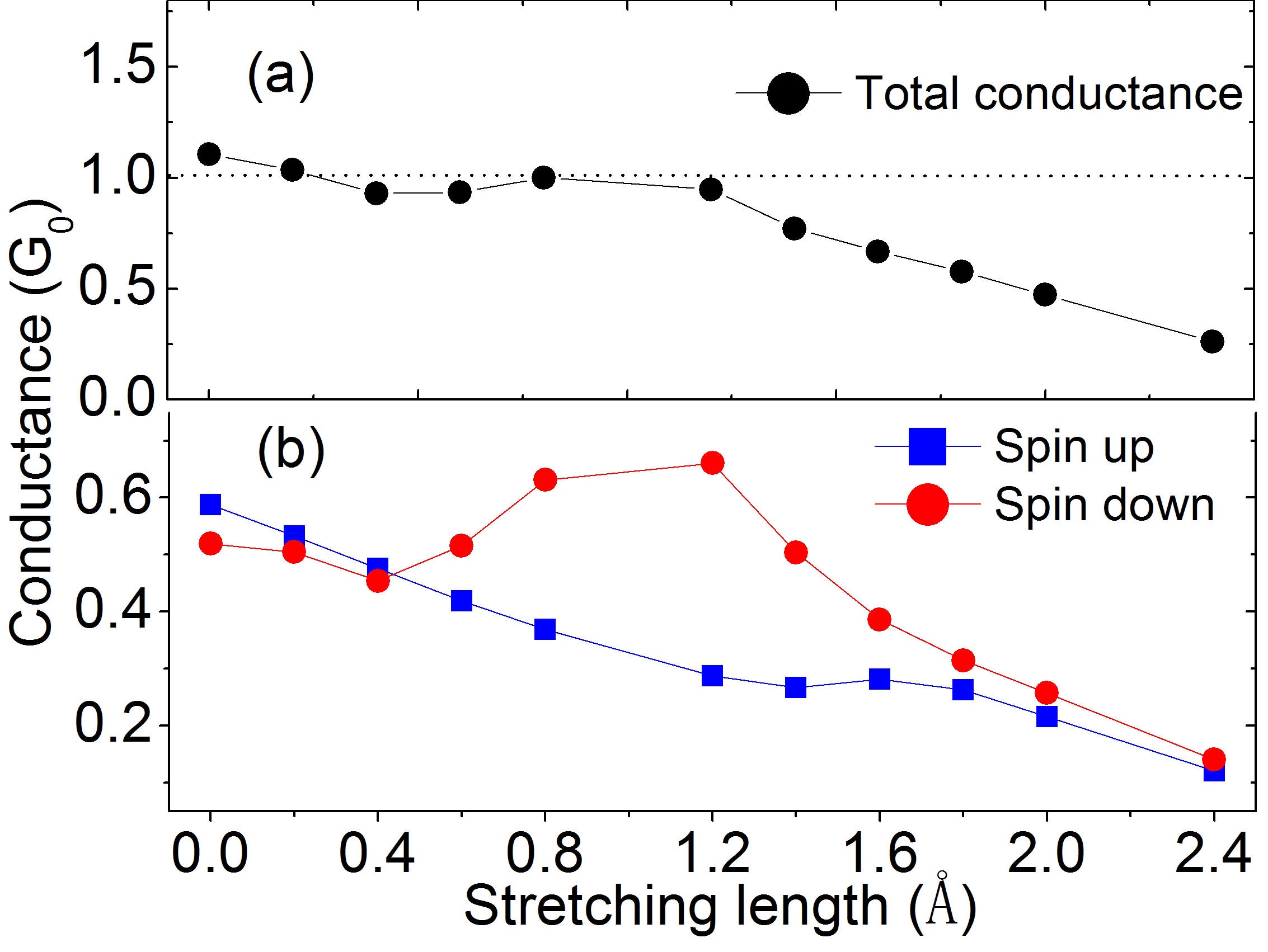}
\caption{\label{fig5}
(a) Variation of the total conductance of the $Ni-2H-Ni$ junction with the stretching length, and (b) the spin polarized conductance.}
\end{center}
\end{figure}

The variation of the conductance with the stretching length is shown in Fig. \ref{fig5}(a). Noticeably, the value of conductance remains approximately 1 $G_0$ for a long stretching length from 0.2 \AA to 1.2 \AA, after which the conductance decreases sharply as the junction is further stretched. The spin-polarized conductance is shown in Fig. \ref{fig5}(b). For a stretching length less than 0.4 \AA, the conductance has a main contribution from the spin-up channels, while above 0.4 \AA the spin-down channels play a dominant role.
 
\begin{figure}[h]
\begin{center}
\includegraphics[scale=0.15]{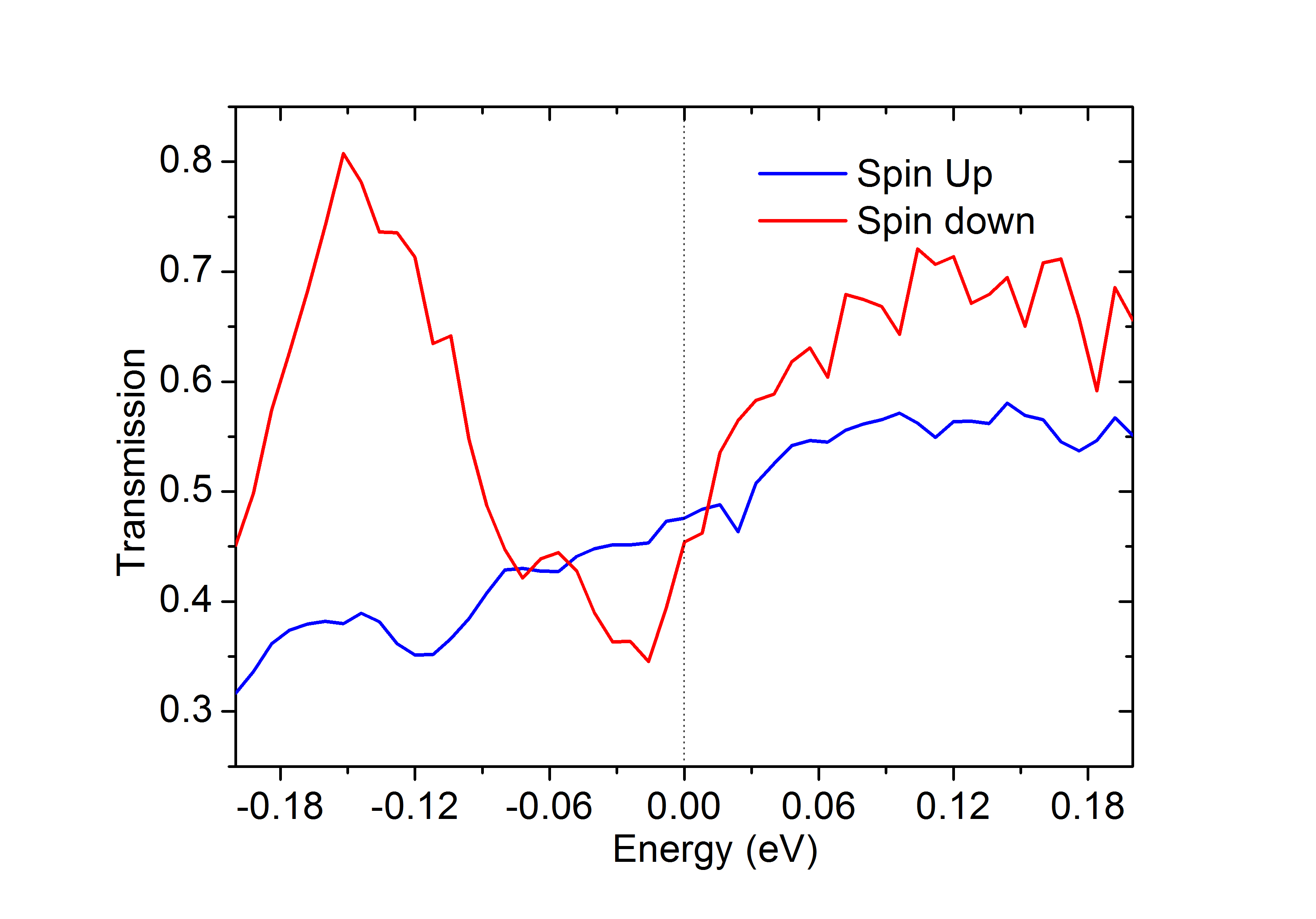}
\caption{\label{fig6}
Transmission of the $Ni-2H-Ni$ junction at the stretching length of 0.4 \AA. Zero energy point is the Fermi energy.}
\end{center}
\end{figure}

Moreover, the experimentally measured Fano Factors for the Ni junction have suggested that the 1 $G_0$ conductance would be weakly polarized \cite{prb-91-245404}. This argument is supported by the calculations done in this study. It was found that the spin polarization is only about 0.2\% at the stretching lengths of 0.2 and 0.4 \AA. The transmission spectrum at 0.4 \AA is plotted in Fig. \ref{fig6}. It shows that there is little difference between the values of spin-up and spin-down conductance at the Fermi energy, indicating a weak spin polarization. From the above analysis, it was concluded that the Ni junction bridged by the two single H atoms could produce the weakly spin-polarized conductance of approximately 1 $G_0$, which explains the experimental observations \cite{prb-91-245404}.

It is worth noting that a similar configuration also accounts for the 1 $G_0$ conductance for the Co \cite{jpcl-1-923} and Fe \cite{prb-69-081401} magnetic junctions. This suggests that for a magnetic metal atomic junction, the absorption of the two separated H atoms is a possible origin of the 1 $G_0$ conductance, while for the nonmagnetic metal junction such as the Rh \cite{jap-117-064310} and Pt \cite{prl-94-036807} junctions, the absorption of a $H_2$ molecule can contribute to the 1 $G_0$ conductance.

\begin{figure}[h]
\begin{center}
\includegraphics[scale=0.6]{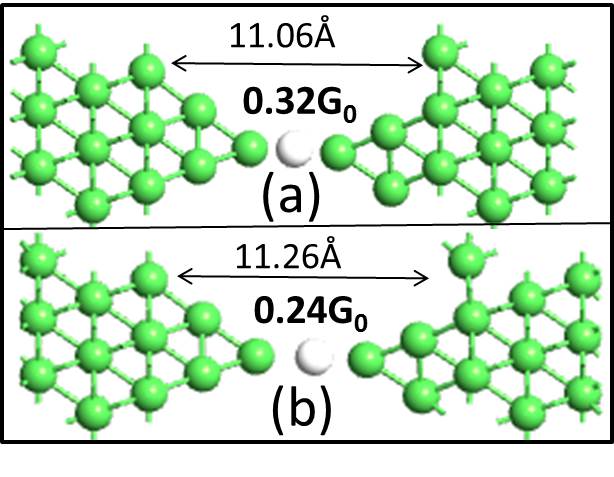}
\caption{\label{fig7}
Conductance of the $Ni-H-Ni$ junction at different stretching length.}
\end{center}
\end{figure}

In addition, the conductance of the $Ni-H-Ni$ junction bridged by a single H atom was also calculated. The conductance is less than 0.35 $G_0$ as the junction is stretched, as shown in Fig. \ref{fig7}. This means that the $Ni-H-Ni$ junction is not likely to generate a 1 $G_0$ conductance.
 
\section{Conclusions}

In summary, the spin-polarized transport properties of the Ni atomic junction bridged by the $H_2$ molecule or single H atoms, using the NEGF method combined with the first-principles calculations, were studied. The conductance is obtained during the stretching process of the junctions. We found that the Ni junction bridged by the $H_2$ molecule has a conductance about 0.7 $G_0$. In contrast, the Ni junction bridged by the two single H atoms has a conductance around 1 $G_0$, which is weakly spin-polarized. For the Ni junction connected by a single H atom, the conductance is much smaller than 1 $G_0$. The results of this study help to understand the physical origin of the 1 $G_0$ conductance observed in related experiments.

\section{Acknowledgments}

This work was financially supported by the National Natural Science Foundation of China (NSFC) under Grant Nos. 11674231 and 11504395.

\bibliographystyle{unsrt}
\bibliography{najha}

\end{document}